\documentclass[useAMS,usenatbib,usegraphicx]{mn2e}
\usepackage{amssymb}

\title[Power spectrum of GLAST point sources]{Angular power spectrum
of gamma-ray sources for GLAST: blazars and clusters of galaxies}
\author[S. Ando et al.]{Shin'ichiro Ando,$^{1,2}$\thanks{E-mail:
  ando@tapir.caltech.edu} Eiichiro Komatsu,$^{3}$ Takuro Narumoto$^{4}$
  \newauthor
  and Tomonori Totani$^{4}$\\
  $^1$Theoretical Astrophysics, California Institute of Technology,
  Pasadena, CA 91125, USA\\
  $^2$Kellogg Radiation Laboratory, California Institute of Technology,
  Pasadena, CA 91125, USA\\
  $^3$Department of Astronomy, University of Texas at Austin, Austin,
  TX 78712, USA\\
  $^4$Department of Astronomy, School of Science, Kyoto University,
  Kyoto 606-8502, Japan}
\begin{document}

\date{Received 5 October 2006; accepted 18 December 2006}

\pagerange{\pageref{firstpage}--\pageref{lastpage}} \pubyear{2006}

\maketitle

\label{firstpage}

\begin{abstract}%
Blazars, a beamed population of active galactic nuclei, radiate
 high-energy $\gamma$-rays, and thus are a good target for the {\it
 Gamma Ray Large Area Space Telescope (GLAST)}.
As the blazars trace the large-scale structure of the universe, one may
 observe spatial clustering of blazars.
We calculate the angular power spectrum of blazars that would be
 detected by GLAST.
We show that we have the best chance of detecting their clustering at
 large angular scales, $\theta\gtrsim 10\degr$, where shot noise is
 less important, and the dominant contribution to the correlation comes
 from relatively low redshift, $z\lesssim 0.1$.
The GLAST can detect the correlation signal, if the blazars detected by
 GLAST trace the distribution of low-$z$ quasars observed by optical
 galaxy surveys, which have the bias of unity.
If the bias of blazars is greater than 1.5, GLAST will detect the
 correlation signal unambiguously.
We also find that GLAST may detect spatial clustering of clusters of
 galaxies in $\gamma$-rays.
The shape of the angular power spectrum is different for blazars and
 clusters of galaxies; thus, we can separate these two contributions on
 the basis of the shape of the power spectrum.
\end{abstract}
\section{Introduction}
\label{sec:Introduction}
Active galactic nuclei (AGNs) are highly energetic astrophysical
objects, which are often accompanied with relativistic jets powered by
accretion onto supermassive black holes located at the central region of
galaxies.
The AGNs radiate in a wide frequency range, from radio waves to
$\gamma$-rays.
The Energetic Gamma Ray Experiment Telescope (EGRET) on board the {\it
Compton Gamma Ray Observatory (CGRO)}, which detects $\gamma$-rays in
GeV energy, have found $\sim 60$ AGNs with its all-sky survey campaign,
and all of them but one (M87) were classified as ``blazars''
\citep{Hartman1999}.
The features in the spectrum and light curves of these blazars indicate
that the relativistic jets are directed towards us.

The {\it Gamma Ray Large Area Space Telescope (GLAST)} is equipped with
the Large Area Telescope (LAT) instrument, which is an upgraded version
of the EGRET.
Its large effective area ($10^4$ cm$^2$) as well as good angular
resolution improve a point-source sensitivity by almost two orders of
magnitude compared to the EGRET, which would increase the source
statistics significantly.
It is expected that a thousand to ten thousand blazars would be detected
as point sources by GLAST \citep{Stecker1996,Chiang1998,Mucke2000,
Narumoto2006}. 
Such a dramatic improvement in point-source sensitivity would allow us
to determine the $\gamma$-ray luminosity function (GLF) of blazars with
unprecedented accuracy.
In contrast, the current blazar GLF is based upon merely $\sim 60$
blazars detected by EGRET.

As the blazars should trace the large-scale structure of the universe,
they should exhibit spatial clustering.
In this paper, we investigate whether the spatial clustering of blazars
is detectable by GLAST, especially by focusing on the angular power
spectrum, a quantity projected along the line of sight.
This would provide the first direct measurement of the bias parameter
of blazars, immediately after GLAST starts taking the data, 
which should be compared
with the results from other classes of AGNs in order to get further insight
into the unification picture of AGNs.
In addition to blazars, it has been pointed out that GLAST may detect
$\gamma$-ray emission from clusters of galaxies
\citep{Colafrancesco1998,Totani2000}.
We thus also calculate the spatial clustering of clusters of galaxies in
$\gamma$-rays and discuss its detectability with GLAST.

This paper is organized as follows.
In Section~\ref{sec:Angular power spectrum of blazars} we calculate the
angular power spectrum of blazars that would be detected by GLAST.
In Section~\ref{sub:Detectability of the blazar correlation} we study
detectability of the spatial correlation of blazars and discuss current
(indirect) observational constraints on the bias of blazars. 
In Section~\ref{sec:Angular power
spectrum of galaxy clusters} we calculate the angular power spectrum of 
galaxy clusters. 
Section~\ref{sec:Discussion} is devoted to further discussions, 
and we conclude in Section~\ref{sec:Conclusions}.
\section{Angular power spectrum of blazars}
\label{sec:Angular power spectrum of blazars}
\subsection{Formalism}
\label{sub:Formulation and results of blazar angular power spectrum}
The angular power spectrum of blazars that would be detected by GLAST is
given by the sum of the shot (Poisson) noise term, $C_l^P$, and the 
correlation term, $C_l^C$, as \citep{Peebles1980}
\begin{eqnarray}
 C_l &=& C_l^P + C_l^C,
  \label{eq:C_l}\\
 C_l^P &=& \mathcal N^{-1},
  \label{eq:C_l^P}\\
 C_l^C &=& 2 \pi \int_{-1}^{1} d\cos\theta\
  P_l(\cos\theta) w(\theta),
  \label{eq:C_l^C}
\end{eqnarray}
where $\mathcal N \equiv dN/d\Omega$ is the number of blazars per solid
angle, and $w(\theta)$ is the angular correlation function of blazars
that would be detected by GLAST.
Note that the shot noise term is independent of multipoles.

A standard procedure to calculate the angular correlation function is as
follows.
We model the 3-d spatial correlation function of blazars, $\xi(r,z)$, as
the correlation function of dark matter particles, multiplied by the
``bias'' factors that depend on physics of formation and evolution of
blazars in dark matter haloes.
We then project the resulting 3-d correlation function on the sky to
calculate the 2-d angular correlation function of blazars, $w(\theta)$.
As the bias factors depend on redshift and luminosity of blazars, we
model $\xi(r,z)$ as $\xi(r;L_{\gamma,1},L_{\gamma,2}|z) = \xi_{\rm
lin}(r,z) b_B(L_{\gamma,1},z) b_B(L_{\gamma,2},z)$, where $r = |\vec x_2
- \vec x_1 |$ is the distance between two blazars, $L_{\gamma,1}$ and
$L_{\gamma,2}$ are their luminosities, and $\xi_{\rm lin}(r,z)$ is the
3-d correlation function of ${\it linear}$ dark matter fluctuations.
As we show in this paper the angular correlation function of blazars may
be detectable only on large scales, and thus the linear correlation
function and the linear bias model would provide a good approximation.
By projecting the 3-d correlation function on the sky, one obtains
\citep{Peebles1980}
\begin{eqnarray}
 \mathcal N^2 w(\theta) &=& \int_0^{z_{\rm max}}
  dz\ \frac{d^2V}{dzd\Omega} \chi(z)^2 \phi (z)^2
  \overline{b}_B(z)^2
  \nonumber\\&&{}\times
  \int_{-\infty}^{\infty} du\
  \xi_{\rm lin} \left(\sqrt{u^2 + \chi(z)^2\theta^2},z\right),
  \label{eq:relation between w and xi}
\end{eqnarray}
where $\chi(z)$ is the comoving distance out to an object at $z$,
$d^2V/dzd\Omega$ is the comoving volume element per unit solid angle and
per unit redshift range, $\overline b_B(z)$ is the average bias of
blazars weighted by the $\gamma$-ray luminosity function (GLF) of
blazars,\footnote{The luminosity function represents the number of
sources per unit comoving volume and unit luminosity range.}
$\rho_\gamma(L_\gamma,z)$:
\begin{equation}
 \overline{b}_B(z) \equiv \frac1{\phi(z)}
  \int_{L_\gamma(F_{\gamma,{\rm lim}},z)}^\infty dL_\gamma\
  \rho_\gamma(L_\gamma,z)  b_B(L_\gamma,z), 
  \label{eq:averaged bias}
\end{equation}
and $\phi(z)$ is the cumulative GLF of blazars, i.e., GLF integrated
from a given minimum luminosity cut-off,
\begin{equation}
 \phi(z) \equiv
  \int_{L_\gamma (F_{\gamma,{\rm lim}},z)}^{\infty}
  dL_\gamma\ \rho_\gamma (L_\gamma,z).
  \label{eq:phi function}
\end{equation}
Note that we have not used the so-called ``small-angle approximation''
or ``Limber's approximation,'' as we are mainly interested in the
signals on large angular scales, $\theta\gtrsim 10\degr$.

We calculate $\xi_{\rm lin}(r,z)$ from the power spectrum of linear
matter density fluctuations, $P_{\rm lin}(k)$:
\begin{equation}
\xi_{\rm lin}(r,z) = \int \frac{k^2 dk}{2\pi^2}\ P_{\rm lin}(k)
\frac{\sin kr}{kr}.
\end{equation} 
We use the linear transfer function given in \citet{Eisenstein1999} to
compute $P_{\rm lin}(k)$.

Equations~(\ref{eq:C_l^P}) and (\ref{eq:C_l^C}) suggest that
$C_l^P={\cal N}^{-1}$ dominates when the number of blazars detected by
GLAST is small, making it difficult to detect the correlation term.
It is therefore very important to understand how many blazars one can
detect with GLAST.
In the next subsection we calculate the expected number count of blazars
for GLAST using the latest GLF of blazars \citep{Narumoto2006}.

\subsection{Gamma-ray luminosity function of blazars}
\label{sec:Gamma-ray luminosity function of blazars}
The basic idea behind the model of the GLF of blazars proposed by
\citet{Narumoto2006} is that the jet activity that powers $\gamma$-ray
emission from blazars must be related to accretion onto the central
black holes, from which X-ray emission emerges; thus, the X-ray and
$\gamma$-ray luminosity of blazars must be correlated.
We use the following relation between GLF of blazars, $\rho_\gamma$, and
X-ray luminosity function (XLF) of AGNs, $\rho_X$:
 \begin{equation}
 \rho_\gamma (L_\gamma,z) = \kappa \frac{L_X}{L_\gamma}
  \rho_X (L_X,z).
  \label{eq:GLF-XLF relation}
\end{equation}
The advantage of this method is that the XLF has been determined
accurately by the extensive study of the X-ray background
\citep{Ueda2003,Hasinger2005}, and thus the predicted GLF would also be
fairly accurate, provided that the $\gamma$-ray luminosity and X-ray
luminosity of blazars are tightly correlated.
Since not all AGNs detected in X-rays are blazars, we have introduced a
normalization factor, $\kappa$.
We relate the $\gamma$-ray luminosity, $L_\gamma$, and X-ray luminosity,
$L_X$, of blazars by a linear relation with the constant of
proportionality given by $10^q$:
\begin{equation}
 L_\gamma = 10^q L_X,
\end{equation}
where $L_\gamma$ represents $\nu L_\nu$ at 100 MeV, and $L_X$ is the
X-ray luminosity integrated over the {\it ROSAT} band, 0.5--2 keV.
(Both are evaluated at the source rest frame.)
We convert the measured flux to the rest-frame luminosity by specifying
the spectrum of sources: for $\gamma$-ray we use an spectral index of
$\alpha_\gamma = 2.2$ \citep{Sreekumar1998}, while for X-ray $\alpha_X =
2$ \citep{Hasinger2005}.

The AGN XLF, $\rho_X$, is given by a double power-law in luminosity with
an evolution factor  $f(L_X,z)$ \citep{Hasinger2005}:
\begin{equation}
 \rho_X (L_X,z) =  
 \frac{A_Xf(L_X,z)}{(\ln 10) L_X}
  \left[\left(\frac{L_X}{L_X^\ast}\right)^{\gamma_1}
  + \left(\frac{L_X}{L_X^\ast}\right)^{\gamma_2}\right]^{-1},
\end{equation}
where
\begin{eqnarray}
 \lefteqn{f(L_X,z)}\nonumber\\&=&
  \left\{
   \begin{array}{lc}
    (1+z)^{p_1} & [z \le z_c(L_X)],\\
    f[L_X,z_c(L_X)]\left[\frac{1+z}{1+z_c(L_X)}\right]^{p_2}
     & [z > z_c(L_X)],
   \end{array}\right.
  \label{eq:XLF evolution function}
\end{eqnarray}
where $z_c$ is the redshift of evolutionary peak given by
\begin{equation}
 z_c(L_X)=
  \left\{
  \begin{array}{lc}
  z_c^\ast & (L_X \ge L_a),\\
  z_c^\ast \left(\frac{L_X}{L_a}\right)^{\alpha} & (L_X < L_a),
  \end{array}
  \right.
  \label{eq:peak redshift}
\end{equation}
and $p_1$ and $p_2$ are given by
\begin{eqnarray}
 p_1 &=& p_1^\ast + \beta_1 \left[\log (L_X/\mathrm{erg\ s^{-1}}) -
 44\right],\\
 p_2 &=& p_2^\ast + \beta_2 \left[\log (L_X/\mathrm{erg\ s^{-1}}) -
 44\right].
 \label{eq:p_1 and p_2}
\end{eqnarray}
\citet{Hasinger2005} have found
 $A_X = 6.69\times 10^{-7}$ Mpc$^{-3}$, $\log
(L_X^\ast / \mathrm{erg\ s^{-1}}) = 43.94\pm 0.11$, $z_c^\ast = 1.96 \pm
 0.15$, $\log (L_a/\mathrm{erg\ s^{-1}}) = 44.67$, $\alpha = 0.21 \pm
 0.04$, $p_1^\ast = 4.7 \pm 0.3$, $p_2^\ast = -1.5 \pm 0.7$, $\beta_1 =
 0.7 \pm 0.3$, $\beta_2 = 0.6 \pm 0.8$, $\gamma_2 = 2.57 \pm 0.16$, and
 $\gamma_1=0.87\pm 0.10$.
We call this model the ``luminosity-dependent density evolution'' model,
LDDE.

How robust are our predictions from this model?
The most important parameter for our purpose in this paper is the slope
of XLF in the faint end, $\gamma_1$, as the expected number count of
blazars that would be detected by GLAST is sensitive to how many blazars
there are in the faint end of luminosity function.
\citet{Narumoto2006} have fitted the GLF of blazars detected by EGRET in
order to find $\gamma_1$, $q=\log (L_\gamma/L_X)$ and $\kappa$, with
the other parameters fixed at the best-fitting values from the XLF given
above.
(The blazar sample from EGRET was constructed by requiring that EGRET
sources were identified as blazars by radio observations. The
probability that the blazars giving the flux above the EGRET point
source sensitivity also gives the sufficient radio flux was taken into
account in their analysis.)
They have found that $(\gamma_1,q,\kappa) = (1.19,3.80,5.11\times
10^{-6})$ best describes the GLF of EGRET blazars.
This $\gamma_1$ is larger than that obtained from the XLF,
$\gamma_1=0.87 \pm 0.10$, at the 3-$\sigma$ level, which may imply that
a better model is needed; however, we do not investigate this point any
further and simply accept $\gamma_1 = 1.19$ as the canonical value for
the GLF of blazars.
One should come back to this point, however, when GLAST flies and
collects many more blazars than available now.

\begin{table}
\caption{Parameters of the LDDE GLFs and the expected number, $N$, and
 the surface number density, ${\cal N}$, of blazars that would be
 detected by GLAST. We have assumed that no blazars fainter than
 $L_{\gamma,{\rm min}} = 10^{41}$~erg~s$^{-1}$ would exist, and GLAST
 LAT can detect the flux down to $F_{\gamma,{\rm lim}} = 2\times
 10^{-9}$ cm$^{-2}$ s$^{-1}$ for 2 years of all-sky observations.}
\label{table:LDDE models}
\begin{minipage}{8.4cm}
\begin{tabular}{lcccc}
 \hline
 Model & $(q,\gamma_1)$ & $\kappa$ & $N$ & $\mathcal N$ (sr$^{-1}$)\\
 \hline
 LDDE1\footnote{Best-fit model of the EGRET blazar distribution.} &
 $(3.80,1.19)$ & $5.11\times 10^{-6}$ & 3100 & 250\\
 LDDE2\footnote{A model explaining 100\% of the EGRB intensity.} &
 $(3.80,1.31)$ & $3.90\times 10^{-6}$ & 6500 & 520\\
 \hline
\end{tabular}
\end{minipage}
\end{table}

One can also calculate the contribution to the extragalactic gamma-ray
background (EGRB) from blazars once the GLF is specified.
The EGRB intensity has been measured by EGRET \citep{Sreekumar1998,
Strong2004},\footnote{But these estimates are still controversial
\citep{Keshet2004b}.} and the best-fit model with $\gamma_1 = 1.19$
accounts for
25--50\% of the EGRB intensity, depending on the assumed minimum
$\gamma$-ray luminosity of blazars, $L_{\gamma,{\rm min}}= 10^{40}$ to
$10^{43}$~erg~s$^{-1}$.
Here, we assume that no blazars fainter than the minimum luminosity
would exist.
On the other hand, blazars can still account for all the EGRB intensity,
if the blazars can be as faint as $L_{\gamma,{\rm min}} =
10^{41}$~erg~s$^{-1}$, and the faint end of the GLF is slightly steeper,
$\gamma_1=1.31$, than the canonical model.
The other parameters are given by $(q,\kappa)= (3.80,3.9 \times
10^{-6}$).
This model appears to be a bit extreme, as $\gamma_1=1.31$ is
inconsistent with the X-ray determination, $\gamma_1=0.87\pm 0.10$, at
the 4.4-$\sigma$ level. Nevertheless, we use this model to show 
the uncertainty in our predictions from the uncertainty in the faint-end
of the GLF.
Henceforth we call the canonical model ($\gamma_1=1.19$) the ``LDDE1''
model, and the latter model ($\gamma_1=1.31$) the ``LDDE2'' model.
For both models, we adopt $L_{\gamma,{\rm min}} = 10^{41}$ erg s$^{-1}$
as the lower luminosity cutoff.

\subsection{Survey parameters, number count, angular correlation
  function, and power spectrum of blazars from GLAST}
\label{sub:Expected number of blazars at the GLAST}

\begin{figure}
\includegraphics[width=8.4cm]{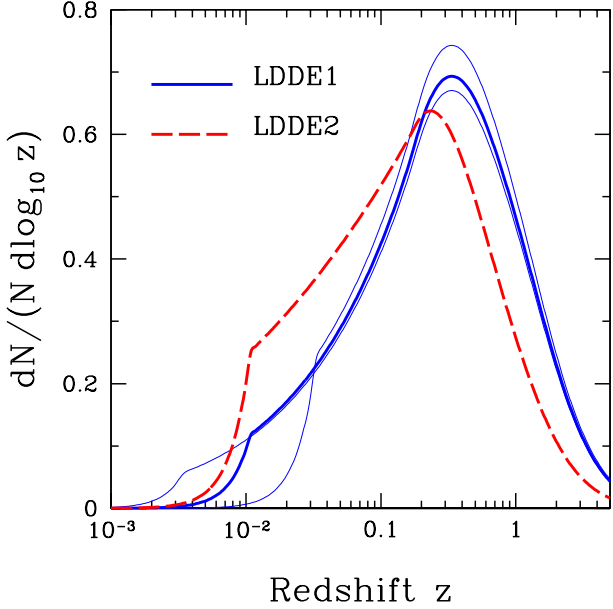}
\caption{Redshift distribution of blazars that would be detected by
 GLAST, for LDDE1 (solid) and LDDE2 (dashed) models. (See
 Table~\ref{table:LDDE models} for the model parameters as well as for
 the expected number of blazars.) The thick solid line shows the LDDE1
 model with $L_{\gamma,\rm min}=10^{41}~{\rm erg~s^{-1}}$, while the
 thin solid lines show the LDDE1 model with $L_{\gamma,\rm
 min}=10^{40}~{\rm erg~s^{-1}}$ and $10^{42}~{\rm erg~s^{-1}}$: the
 larger the $L_{\gamma,\rm min}$ is, the fewer the low-$z$ blazars would
 be detected.}
\label{fig:zdist_blazar}
\end{figure}

The flux sensitivity for point sources of the GLAST LAT is
$F_{\gamma,{\rm lim}} = 2\times 10^{-9}$ cm$^{-2}$ s$^{-1}$ for 2 years
of full-sky observations and for sources having the energy spectral
index of 2; we adopt this value of the flux sensitivity in the following
discussions, unless otherwise stated.
The $\gamma$-ray flux, $F_\gamma$, represents the flux integrated above
$E_{\rm min} = 100$ MeV, and it is related to the $\gamma$-ray
luminosity through
\begin{equation}
 L_{\gamma}(F_\gamma,z) = \frac{4\pi d_L^2(\alpha_\gamma-1)}
  {(1+z)^{2-\alpha_\gamma}} E_{\rm min} F_\gamma,
 \label{eq:limflux}
\end{equation}
where $d_L$ is the luminosity distance.
One can calculate the number of blazars that would be detected by GLAST
from
\begin{equation}
 N = \Omega \int_0^{z_{\rm max}} dz\ \frac{d^2V}{dzd\Omega} \phi(z),
  \label{eq:N}
\end{equation}
where we use $z_{\rm max} = 5$, $\Omega$ is the solid angle covered
($\Omega = 4\pi$~sr for the all-sky survey), and $\phi(z)$ is the
cumulative GLF given by equation~(\ref{eq:phi function}).

For the canonical GLF model that accounts for 25--50\% of the EGRB
intensity (LDDE1) and the lower luminosity cutoff of $L_{\gamma,{\rm
min}} = 10^{41}$~erg~s$^{-1}$ (hereafter, we use this value unless
otherwise stated), we obtain $N \simeq 3100$.
For the GLF model that accounts for all the EGRB intensity (LDDE2) and
the same luminosity cutoff, we obtain $N\simeq 6500$.
These results are summarized in Table~\ref{table:LDDE models}.
Therefore, it is easier to detect the spatial clustering of blazars in
the LDDE2 model than in the LDDE1 model.

Figure~\ref{fig:zdist_blazar} shows the redshift distribution of GLAST
blazars predicted from the LDDE1 and LDDE2 model.
For both cases, the distribution exhibits a sharp cut-off around $z =
0.01$, which is due to our assumption that no blazars fainter than
$L_{\gamma,{\rm min}}$ would exist; a larger $L_{\gamma,{\rm min}}$
results in a larger cut-off redshift.
Nevertheless, since only a small fraction of the distribution is
eliminated  by this effect, the total number of blazars that would be
detected by GLAST, $N$, hardly changes; for example, we expect  3200 and
2900 blazars to be observed by GLAST for $L_{\gamma,{\rm min}} =
10^{40}$ and $10^{42}$ erg s$^{-1}$ (both for the LDDE1 parameters),
respectively.
On the other hand, we shall show in Section~\ref{sub:Signal-to-noise vs
blazar bias} that $L_{\gamma,{\rm min}}$ has an important consequence
for detectability of the anisotropy signal. 

\begin{figure*}
\begin{center}
\includegraphics[height=12cm]{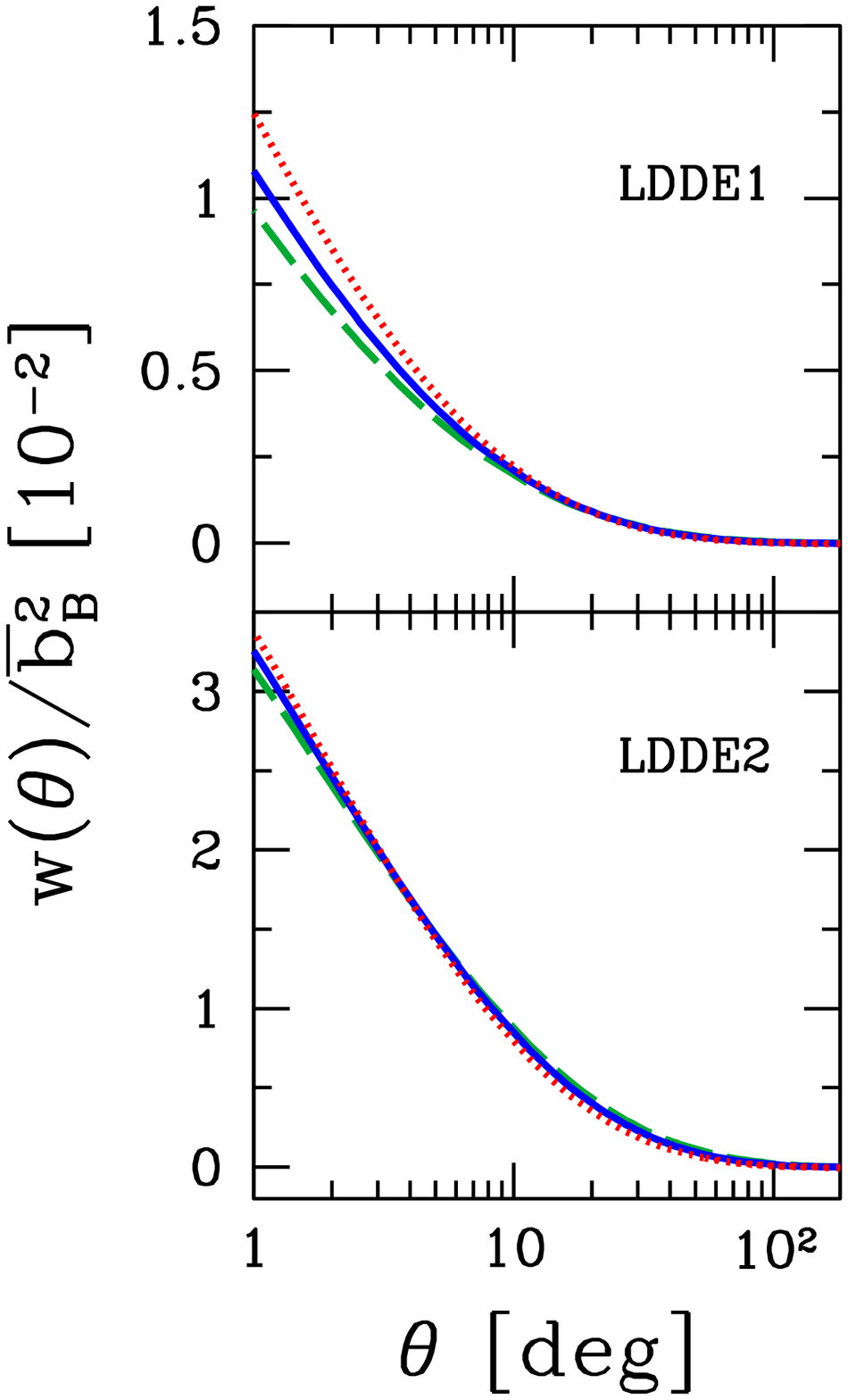}
\hspace{0.5cm}
\includegraphics[height=12cm]{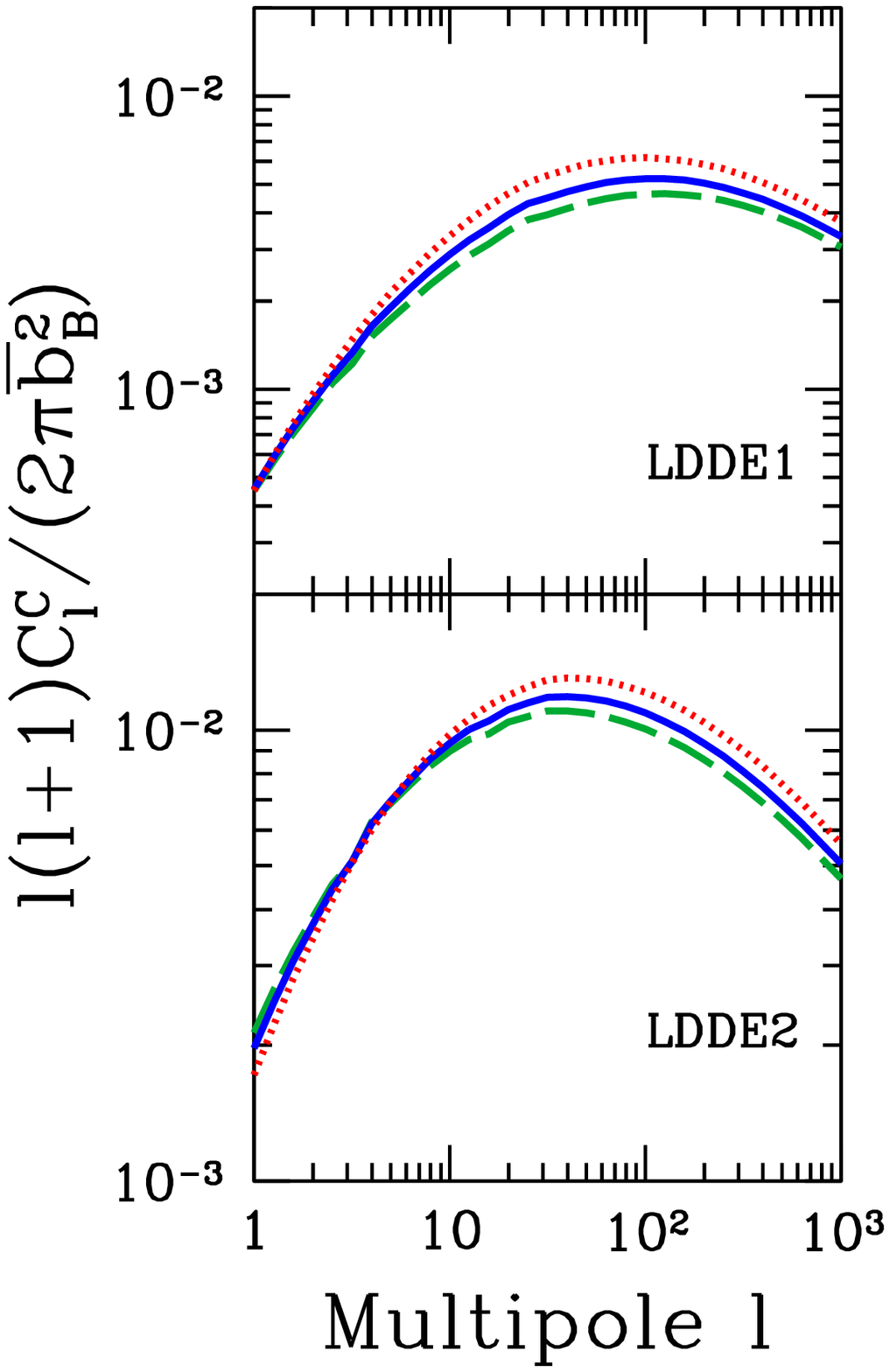}
\caption{({\it Left}) Angular correlation function, $w(\theta)$, and
 ({\it Right}) correlation term of the angular power spectrum,
 $l(l+1)C_l^C/2\pi$, of the blazars that would be detected by
 GLAST. Both have been divided by the average bias squared; thus plotted
 quantities are $w(\theta)/\overline{b}^2_B$ and
 $l(l+1)C_l^C/(2\pi\overline{b}^2_B)$. The dotted, solid, and dashed
 lines show the predictions for the limiting flux of $F_{\gamma,{\rm
 lim}}= 2$, 3, and $4\times 10^{-9}~{\rm cm^{-2}~s^{-1}}$,
 respectively. The top panels are for the LDDE1 model, while the bottom
 panels are for the LDDE2 model.}
\label{fig:summary}
\end{center}
\end{figure*}

Figure~\ref{fig:summary} shows the angular correlation function,
$w(\theta)$ (left panels), and the correlation term of the angular power
spectrum, $l(l+1)C^C_l/2\pi$ (right panels), divided by the average bias
squared, for the LDDE1 (top panels) and LDDE2 (bottom panels) model.
In each panel we vary the GLAST LAT point-source flux sensitivity,
$F_{\gamma,{\rm lim}}$, from $2\times 10^{-9}$ to $4\times 10^{-9}~{\rm
cm^{-2}~s^{-1}}$.
As expected, the clustering is stronger when more sources are observed,
i.e., LDDE2 and lower $F_{\gamma,{\rm lim}}$.

\section{Detectability of the blazar correlation}
\label{sub:Detectability of the blazar correlation}

\begin{figure*}
\begin{center}
\rotatebox{-90}{\includegraphics[width=12cm]{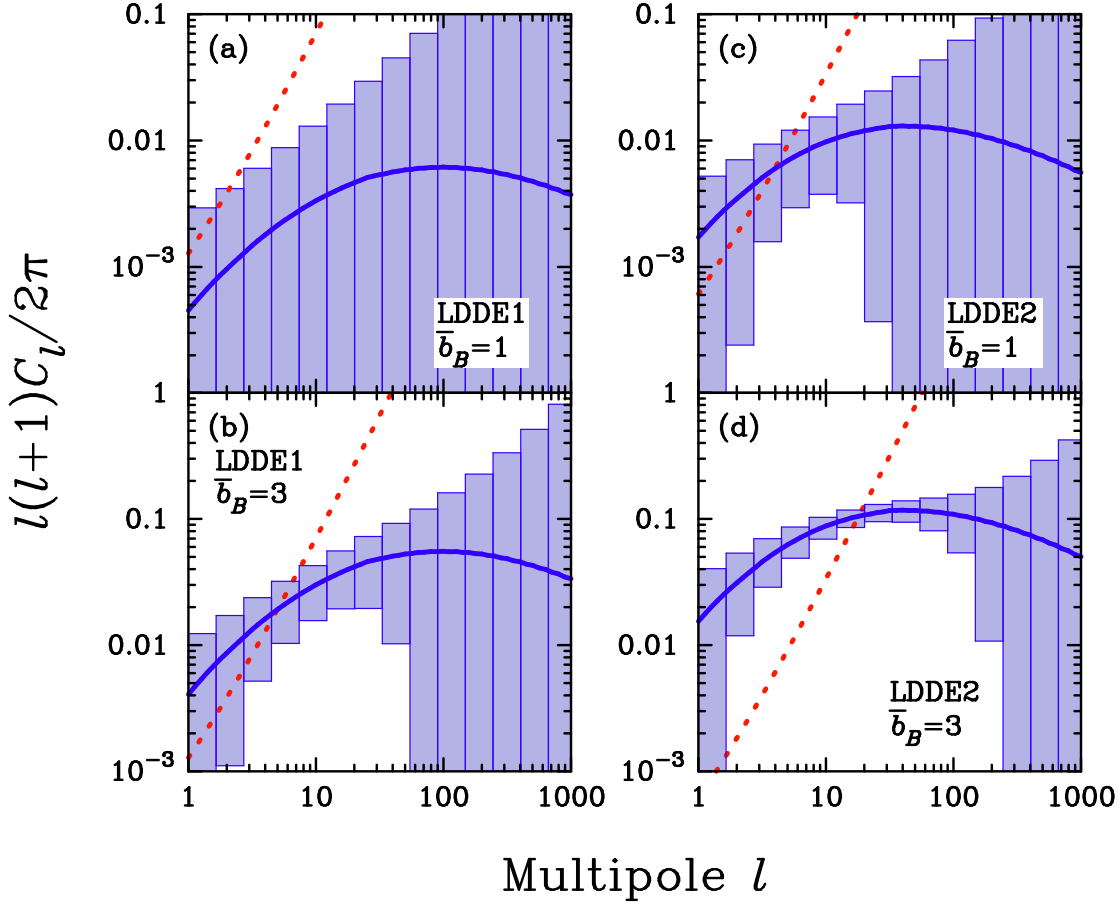}}
\caption{Angular power spectrum of GLAST blazars, $l(l+1)C_l/2\pi$.
 The dotted lines show the shot noise term, $C_l^P$
 [equation~(\ref{eq:C_l^P})], while the thick solid lines show the
 correlation term, $C_l^C$ [equation~(\ref{eq:C_l^C})], for the following
 models: (a) LDDE1,  $\overline b_B = 1$, (b) LDDE1,  $\overline b_B =
 3$, (c) LDDE2,  $\overline b_B = 1$, and (d) LDDE2,  $\overline b_B = 3$.
 The boxes show the 1-$\sigma$ errors in $C_l^C$ binned with $\Delta l =
 0.5 l$ [equation~(\ref{eq:delta C_l})].}
\label{fig:C_l_blazar}
\end{center}
\end{figure*}

\subsection{Signal-to-noise vs blazar bias}
\label{sub:Signal-to-noise vs blazar bias}

As the correlation function and power spectrum are proportional to
the average bias squared, $w(\theta)\propto \overline{b}_B^2$ and
$C_l^C\propto \overline{b}_B^2$, whether or not one can detect the
angular clustering of blazars crucially depends on $\overline{b}_B$.
Before we investigate a model of the blazar bias, let us ask this
question, ``how large $\overline{b}_B$ should be, in order for $C_l^C$
to be detected by GLAST?''

The statistical error in the measurement of $C_l$ is given by the
following argument.
Assuming statistical isotropy of the universe, we have $2l+1$
independent samples of $C_l=|a_{lm}|^2$ (with different $m$'s) per
multipole.
Here, $a_{lm}$ is the spherical harmonic coefficient of the distribution
of blazars on the sky.
One may thus estimate $C_l$ from $C_l=\sum_{m=-l}^l|a_{lm}|^2/(2l+1)$.
The error in $C_l$ is given by
\begin{equation}
 (\delta C_l)^2 = \frac{2C_l^2}{(2 l + 1)\Delta l f_{\rm sky}}
= \frac{2(C_l^P+C_l^C)^2}{(2 l + 1)\Delta l f_{\rm sky}}
  \label{eq:delta C_l}
\end{equation}
where $\Delta l$ is the bin size in $l$ space and $f_{\rm sky}$ is a
fraction of the sky covered by observations.
For the all-sky survey like GLAST, we may adopt $f_{\rm sky} = 1$; we
note that the point source sensitivity becomes worse near the galactic
plane because of strong galactic foreground.
As $C_l^P={\cal N}^{-1}$ is independent of $l$ and depends only on the
inverse of the surface density of blazars, one can fit it and subtract
it from the measured $C_l$, leaving only $C_l^C$.
The error in $C_l^C$, however, still contains the contribution from
$C_l^P$.
This shows why it is important to detect as many blazars as possible
(and thus reduce $C_l^P$ as much as possible), in order to measure
$C_l^C$.

Figure~\ref{fig:C_l_blazar} shows the 1-$\sigma$ error boxes binned with
$\Delta l = 0.5 l$ for the LDDE1 and LDDE2 model.
We show the errors for the average bias of $\overline{b}_B=1$ and 3.
(Note that we have ignored the redshift dependence of $\overline{b}_B$.)
We find that it would be difficult to detect $C_l^C$ for
the LDDE1 plus $\overline{b}_B=1$ model, while the other models yield
sufficient signal-to-noise ratios. 

\begin{figure}
\begin{center}
\includegraphics[width=8.4cm]{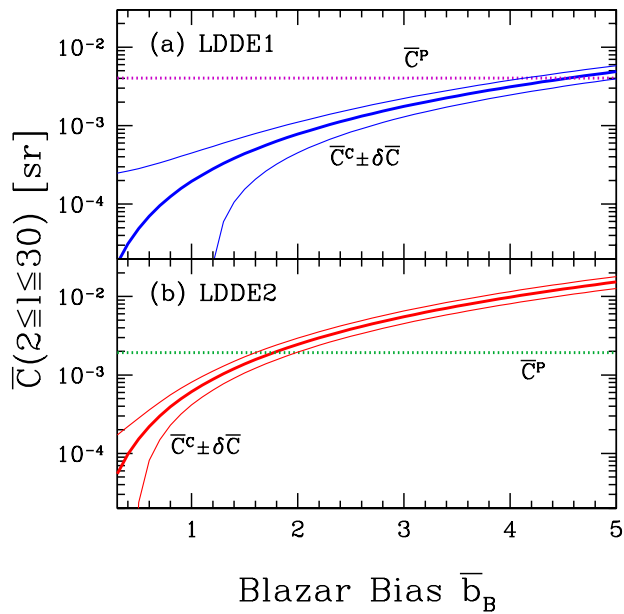}
\caption{Angular power spectrum averaged over $2\leq l\leq 30$, 
 $\overline C$ [equation~(\ref{eq:averaged C_l})], as a function of the
 average bias of blazars. The dotted lines show the Poisson term,
 $\overline C^P$, while the thick solid lines show the correlation term,
 $\overline C^C$, for (a) LDDE1, and (b) LDDE2. The thin solid lines
 show the 1-$\sigma$ errors in $\overline C^C$ [equation~(\ref{eq:error of
 averaged C_l})].}
\label{fig:C_l_bias}
\end{center}
\end{figure}

To increase statistical power one may sum $C_l$ over multipoles.
Let us define the angular power spectrum averaged over $2\le l \le
30$,\footnote{A dipole component, $C_1$, depends on Earth's motion and
is not considered here.}
\begin{equation}
 \overline C(2\le l\le 30) = \frac{1}{29} \sum_{l=2}^{30} C_l.
  \label{eq:averaged C_l}
\end{equation}
The errors of this quantity is then given by
\begin{eqnarray}
 (\delta \overline C)^2 &=&
  \sum_{l=2}^{30}
  \left(\frac{\partial \overline C}{\partial C_l}\right)^2
  \left[\delta C_l(\Delta l = 1)\right]^2
  \nonumber\\&=&
  \frac{1}{29^2}\sum_{l=2}^{30}\frac{2}{(2l+1)f_{\rm sky}}(C_l^P+C_l^C)^2.
  \label{eq:error of averaged C_l}
\end{eqnarray}
Figure~\ref{fig:C_l_bias} shows $\overline C^C(2\le l\le 30)$ as a
function of the average blazar bias, $\overline b_B$, for the LDDE1 (top
panel) and LDDE2 (bottom panel) models.
The expected 1-$\sigma$ errors as well as the Poisson contribution,
$\overline C^P$, are also shown.
For the LDDE1 model we find that GLAST can detect $\overline{C}^C$ if
$\overline{b}_B\gtrsim 1.2$.
For the LDDE2 model the detection is much easier, even for
$\overline{b}_B\gtrsim 0.5$.

Our results depend on the luminosity cutoff of the GLF, $L_{\gamma,{\rm
min}}$, as the correlation at large separations ($l \lesssim 30$) is
dominated mainly by relatively nearby (less bright) sources.
We have therefore performed the same calculations with  different
$L_{\gamma,{\rm min}}$  (with the other parameters of the LDDE1 model
held fixed), and found that the correlation would be detectable (i.e.,
$\overline C^C / \delta \overline C > 1$) for the average bias greater
than 0.9 and 1.7, for $L_{\gamma,{\rm min}} = 10^{40}$ and $10^{42} \
\mathrm{erg \ s^{-1}}$, respectively.

One may also ask how these results would change, if we chose other GLF
models.
The ``pure-luminosity evolution'' (PLE) model has been used
traditionally in the literature \citep{Stecker1996,Chiang1998}, while
the LDDE model fits the EGRET blazar properties better
\citep{Narumoto2006}.
Motivated by the correlation between radio and $\gamma$-ray luminosities
of blazars, \citet{Stecker1996} used the PLE model to obtain the GLF of
blazars.
We find that the large-angle correlation ($l \le 30$) is more difficult
to detect in their model: the correlation would be detectable only when
$\overline b_B > 4.2$.
Their model, however, was not intended to reproduce the redshift and
luminosity distributions of the EGRET blazars, and thus their fit to
these data is not very good.
\citet{Chiang1998} improved the PLE model by adjusting a few parameters
such that the model can reproduce the distribution of EGRET blazars.
(Although the authors did not use the radio and $\gamma$-ray luminosity
relation, we incorporate this in our calculations; see
\citet{Narumoto2006} for details.)
Again, we find that the correlation signal is more difficult to detect
in the best-fit PLE model: the correlation would be detectable only when
$\overline b_B > 6.9$.
These results are because the PLE model predicts the blazar distribution
that is much more biased toward the high redshift (see Fig.~11 of
\citet{Narumoto2006}), and hence, the large-separation power (due mainly
to low-redshift blazars) is suppressed.
In fact, the results improve if we instead adopt the smaller separation,
$30 \le l \le 300$, where the high-redshift contribution becomes
larger.
The sensitivity to the bias parameter goes down to $\overline b_B > 2.4$
and $\overline b_B > 3.0$, respectively for the \citet{Stecker1996} and
\citet{Chiang1998} models.
On the other hand, as we have already shown, the latest GLF from the
LDDE model, which best describes the distribution of EGRET blazars,
predicts that the correlation would be detectable for $\overline b_B$ of
order unity.

\subsection{Modeling blazar bias}
\label{sec:Modeling blazar bias}
The bias of blazars is not known, and thus GLAST may provide the first
determination of the blazar bias, if it is greater than 1.2 (for LDDE1;
0.5 for LDDE2).
In this section we estimate the blazar bias from the existing
observations using three different (indirect) techniques: the bias of
quasars, X-ray observations, and halo model.
However, we emphasize that none of these estimates can be very accurate
(and in fact, they disagree each other)---we will need GLAST to give us
the answer.

\subsubsection{Bias from optical quasar surveys}
\label{sub:Bias from optical quasar surveys}
If the blazar is truly a beamed population of AGNs, then its bias should
be correlated with that of AGNs observed in other wavebands.
It is therefore natural to use such information to estimate the bias of 
blazars.

\begin{figure}
\begin{center}
\includegraphics[width=8.4cm]{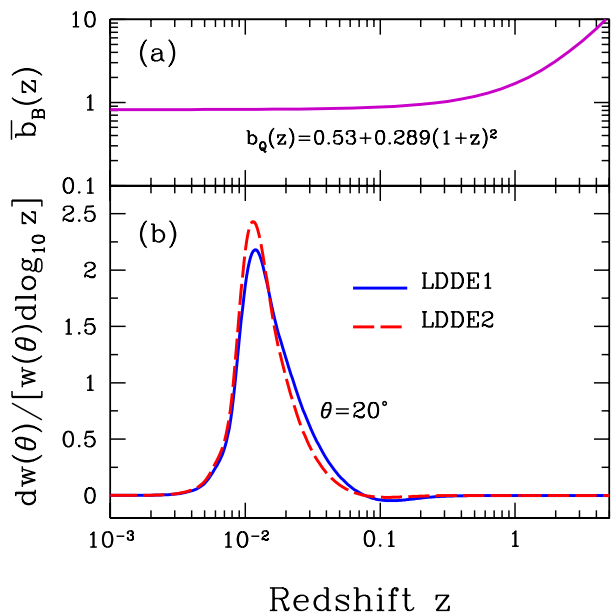}
\caption{(a) Quasar bias, $b_Q(z)$, from the Two-degree Field Quasar
 Redshift Survey, and (b) contribution to the angular correlation
 function per $\ln z$, $dw(\theta)/[w(\theta)d\ln z]$, at $\theta =
 20\degr$, for the LDDE1 (solid) and LDDE2 (dashed) models. We have used
 $b_Q(z)$ for the average blazar bias, $\overline b_B(z)$.}
\label{fig:bias_QSO}
\end{center}
\end{figure}

The optical quasar surveys, such as the Two-degree Field Quasar Redshift
Survey (2QZ) and the Sloan Digital Sky Survey (SDSS), may be the most
efficient way for doing this.
Figure~\ref{fig:bias_QSO}(a) shows the spectroscopic result that the 2QZ
suggests \citep{Croom2005}
\begin{equation}
 b_Q(z) = 0.56 + 0.289 (1+z)^2,
  \label{eq:quasar bias}
\end{equation}
which is also consistent with the photometric result from the SDSS
\citep{Myers2006}.
By comparing $b_Q(z)$ with the bias of dark matter haloes (which is
discussed later), \citet{Croom2005} found that the mass of dark matter
haloes that host quasars is typically $\sim 4\times 10^{12} M_{\sun}$,
almost independent of redshift.

\begin{figure}
\begin{center}
\includegraphics[width=8.4cm]{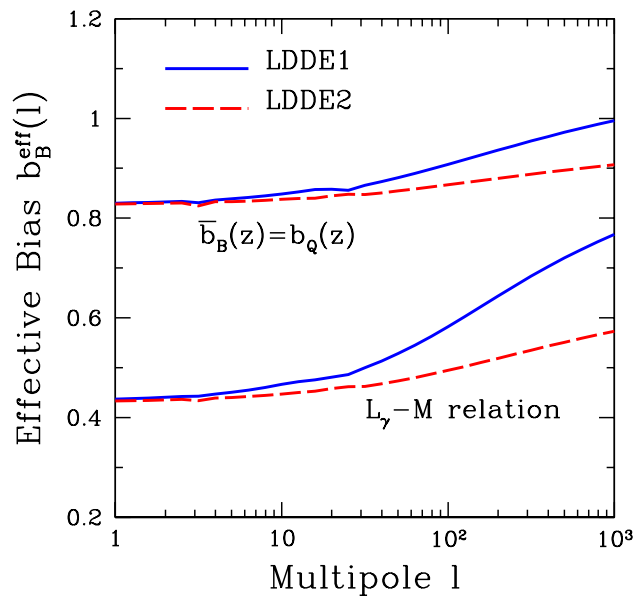}
\caption{The effective bias, $b^{\rm eff}_B(l)$, for the LDDE1 (solid)
 and LDDE2 (dashed) models. The top curves are from $\overline b_B(z) =
 b_Q(z)$, while the bottom curves are from the halo model with
 $L_\gamma$--$M$ relation [equation~(\ref{eq:L-M relation})].}
\label{fig:bias_eff}
\end{center}
\end{figure}

In order to make a quantitative comparison between the results obtained
here and those in the previous section, we define an ``effective bias,''
$b^{\rm eff}_B(l)$, by
\begin{equation}
 b^{\rm eff}_B(l) = \sqrt{
  \frac{C_l(\mbox{model bias, } \overline b_B(z))}
  {C_l(\overline b_B=1)}},
  \label{eq:effective bias}
\end{equation}
where the numerator is $C_l^C$ calculated with $\overline b_B(z) =
b_Q(z)$, and the denominator is that calculated with $\overline b_B =
1$.
If $b^{\rm eff}_B(l)$ is greater than 1.2 and 0.5 for the LDDE1 and
LDDE2 models, respectively, GLAST can detect $C_l^C$ from this
population.
The top curves of Fig.~\ref{fig:bias_eff} show $b^{\rm eff}_B(l)$ as a
function of $l$.
We find that $b^{\rm eff}_B(l)\sim 0.8$ for a relevant range of
multipoles; thus, $C_l^C$ is detectable for the LDDE2 model but not for
the LDDE1 model.
We also find that $b^{\rm eff}_B(l)$ increases as $l$ does (haloes at
higher redshifts are contributing more to the small angular scales),
although the dependence is only modest and is not able to bring the bias
to high enough values for detection for the LDDE1 model.

Since the multipole $l$ is roughly related to the angular separation via
$\theta \approx 180\degr / l$, the angular power spectrum at $l \sim
10$ contributes most to the angular correlation function, $w(\theta)$,
at $\theta \approx 20\degr$.
In Fig.~\ref{fig:bias_QSO}(b), we show that the contribution to
$w(\theta)$ per $\ln z$ peaks at $z\sim 0.01$ with a tail extending up
to $z\sim 0.1$.
At such low redshifts the value of the averaged bias is $\sim 0.8$,
almost independent of $z$.

\subsubsection{Bias from X-ray point-source surveys}
\label{sub:Bias from X-ray point-source surveys} 
The GLF proposed by \citet{Narumoto2006} that we are using in this paper
was derived on the basis of a correlation between $\gamma$-ray and X-ray
luminosities of emission from blazars.
Therefore, the bias derived from the spatial clustering of AGNs detected
in X-ray surveys may provide a useful information regarding the bias of
blazars.

Nevertheless, the clustering of AGNs determined from X-ray surveys are
somewhat controversial.
While both the angular and 3-d correlation function of the X-ray bright
AGNs detected by the {\it ROSAT} surveys are consistent with those from
the optical quasar surveys \citep{Vikhlinin1995,Mullis2004}, those from
{\it Chandra} and {\it XMM-Newton} suggest that these sources are
clustered more strongly than optically selected quasars
\citep{Yang2003,Basilakos2005,Gandhi2006}.
The latest determination of $w(\theta)$ from the {\it XMM-Newton} Large
Scale Structure Survey gave $w(30\arcsec) \sim 0.2$ for $\sim 1130$
sources detected over 4.2~deg$^2$ on the sky in the 0.5--2~keV band
\citep{Gandhi2006}, although statistical significance is only $\sim
2\sigma$.
This result may be interpreted as the bias being $\sim 3.7$ (or
$\lesssim 3.7$ at the 2-$\sigma$ level), which we have obtained as
follows: using the LDDE XLF and the best-fitting parameters in
\citet{Hasinger2005}, we have calculated $w(\theta)$ from
equation~(\ref{eq:relation between w and xi}), in which we have used the
spatial correlation function of {\it non-linear} dark matter
fluctuations, as non-linearity cannot be ignored at such a small angular
separation.
(We have used the halo model approach \citep{Seljak2000} to obtain the
non-linear power spectrum.)
If such a large bias is realized also for blazars, then GLAST should
quite easily be able to detect the correlation.
We should, however, keep in mind that no significant correlations were
observed in the 2--10~keV band, and one needs to improve the source
statistics before making any definitive conclusion about the bias of
AGNs from X-ray surveys.

\subsubsection{Bias from halo model}
\label{sub:Blazar luminosity and host-halo mass relation}
Finally, we try to estimate the bias of blazars from the halo model.
As any galaxies (hosting blazars) must form in dark matter haloes, the
bias of blazars should be related to the bias of dark matter haloes,
$b_h(M,z)$, which is known accurately from $N$-body simulations as well
as from analytical models such as the extended Press-Schechter model
\citep[e.g.,][]{Mo1996}.
Let us suppose that the $\gamma$-ray luminosity of blazars is correlated
with the mass of host dark matter haloes, $L_\gamma=L_\gamma(M_h)$.
Then, the bias of blazars may be estimated from the bias of dark matter
haloes via $b_B(L_\gamma,z)=b_h(M_h(L_\gamma),z)$. 
We use $b_h(M_h(L_\gamma),z)$ derived by \citet{Mo1996}.

The relation between the $\gamma$-ray luminosity of blazars and the host
halo mass is not known.
Whether or not such a relation actually exists is not known either.
Moreover, even if there is a relation between $L_\gamma$ and $M_h$, not
all haloes would host blazars, especially when one takes into account
the fact that the jets from blazars should be directed towards us.
We nevertheless estimate $b_B$ here using the following argument.
Since it is plausible that the blazar $\gamma$-rays are emitted via the
inverse-Compton scattering of the relativistic electrons accelerated in
blazar jets off the cosmic microwave background (CMB), the $\gamma$-ray
luminosity would be correlated with the activity of a central
supermassive black hole.
\citet{Wang2002} have found an empirical relation between the blazar
luminosity at the peak frequency (i.e., a frequency at which the blazar
emits most energy, $\nu L_\nu$), $L_{\rm pk}$, in the optical to X-ray
regime, and the luminosity of emission lines, $L_{\rm lines}$, in the
optical.
The line luminosity is related to the Eddington luminosity via $\log
(L_{\rm lines}/L_{\rm Edd}) \approx -5$ to $-3$ \citep{Wang2002} which,
in turn, would give a $L_{\rm pk}$--$M_{\rm BH}$ relation, where $M_{\rm
BH}$ is a black hole mass which determines the Eddington luminosity.
We then relate the $\gamma$-ray luminosity, $L_\gamma$, with the peak
luminosity, $L_{\rm pk}$, by assuming that they are roughly equal,
$L_\gamma \approx L_{\rm pk}$ \citep{Inoue1996,Kataoka1999}.
We still need to relate the black hole mass, $M_{\rm BH}$, with the host
halo mass, $M_h$, for which we use a correlation between the black hole
mass and the host {\it galaxy} mass, $M$ \citep{Ferrarese2006}.
Of course, the galaxy mass may not be equal to the dark halo mass, the
latter being larger, as dark matter haloes extend more than the luminous
part of galaxies.
We find that this uncertainty hardly affects our prediction for the
blazar bias.
Using these arguments, we finally obtain the desired relation,
\begin{equation}
 M = 10^{11.3} M_{\sun}
  \left(
  \frac{L_\gamma}{10^{44.7}\ \mathrm{erg\ s^{-1}}}
  \right)^{1.7},
  \label{eq:L-M relation}
\end{equation}
for $\log (L_{\rm lines}/L_{\rm Edd})=-4$.
We shall use this relation for estimating the blazar bias from
$b_B(L_\gamma,z) = b_h(M(L_\gamma),z)$.

Before we proceed further, let us address the uncertainty in our
treatment.
If we used $\log (L_{\rm lines} / L_{\rm Edd}) = -3$ instead of $-4$,
and  $L_\gamma = 0.1 L_{\rm pk}$ instead of $L_\gamma = L_{\rm pk}$, we
would obtain $M = 10^{13} M_{\sun} (L_\gamma / 10^{44} \ \mathrm{erg\
s^{-1}})^{1.7}$, making the host haloes more massive and hence the
larger bias for the same luminosity.
We find, however, that the resulting $\overline{b}_B$ did not change
significantly.

\begin{figure}
\begin{center}
\includegraphics[width=8.4cm]{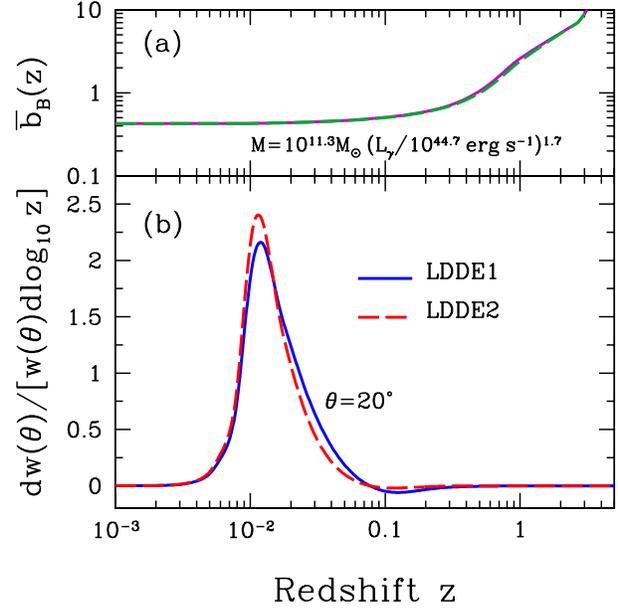}
\caption{The same as Fig.~\ref{fig:bias_QSO}, but for the blazar bias
 equal to the bias of dark matter halos with an empirical $\gamma$-ray
 luminosity--mass relation [equation~(\ref{eq:L-M relation})].}
\label{fig:bias_LM}
\end{center}
\end{figure}

Figure~\ref{fig:bias_LM}(a) shows $\overline b_B(z)$ from
equation~(\ref{eq:averaged bias}) with $b_B(L_\gamma,z) =
b_h(M(L_\gamma),z)$ and equation~(\ref{eq:L-M relation}).
Figure~\ref{fig:bias_LM}(b) shows that the most dominant contribution to
$w(\theta)$ at $\theta=20^\circ$ comes from $z\lesssim 0.1$, which
implies that the angular power spectrum of blazars that would be
detected by GLAST would have the average bias of about 0.4, which is too
small for GLAST to measure $C_l^C$.
Both LDDE1 and LDDE2 give very similar results.
The reason why we found such a small bias is that our model predicts
that GLAST would detect fainter blazars at low redshifts than bright
galaxies at high redshifts.
Therefore, the average bias is dominated by the faint, low-$z$ blazars
that have a small bias.
(Faint blazars are formed in low-mass dark matter halos, which have a
small bias.)
At $z \gtrsim 0.5$, on the other hand, $\overline b_B(z)$ can be as
larger as unity, as only the bright blazars (in massive haloes) are
detectable.

Although it does not affect our analysis very much, let us mention one
subtle feature of the halo approach we just described.
The GLF of blazars may be related to the mass function of dark matter
halos, $dn_h/dM_h$, as
\begin{equation}
 dL_\gamma\ \rho_\gamma (L_\gamma,z)
  = dM_h \ \frac{dn_h}{dM_h}(M_h,z) N_B(M_h,z),
  \label{eq:GLF and mass function}
\end{equation}
where $N_B(M_h,z)$ is the so-called ``halo occupation distribution,''
which represents the average number of blazars per each halo of mass,
$M_h$, at a given redshift, $z$.
One can use this relation, GLF ($\rho_\gamma$), and  $L_\gamma$--$M_h$
relation to obtain  $N_B(M_h,z)$.
Using  the Press-Schechter function \citep{Press1974} for $dn_h/dM_h$
and the LDDE1 model for $\rho_\gamma$, we have found that $N_B(M,z)$
diverges exponentially at the high-mass end.
This is because $dn_h/dM_h$ has an exponential cutoff, while
$\rho_\gamma$ given by the LDDE1 model decreases only as a power law
with the luminosity.
Of course this divergence is an artifact from the fact that we do not
know the precise shape of the GLF at the brightest end, which is poorly
constrained.
It is likely that (i) the GLF has a maximum luminosity above which it
rapidly approaches zero, and (ii) the assumed $L_\gamma$--$M$ relation
cannot be extrapolated to large luminosities.
In order to remove the divergence, we have put an upper cutoff in the
GLF so that $N_B(M,z)$ never exceeds 1.
For the LDDE1 model the cutoff luminosity is $3 \times 10^{47}$ erg
s$^{-1}$, which keeps the GLF still consistent with the EGRET data
because the contribution to the GLF from such luminous blazars is not
significant (see Fig.~12 of \citet{Narumoto2006}).

Any of the assumptions we have made in this subsection could be
incorrect.
The GLAST data will provide us with much better idea about the
clustering of blazars, which will enable us to test these assumptions.

\section{Angular power spectrum of galaxy clusters}
\label{sec:Angular power spectrum of galaxy clusters}
\subsection{Gamma-ray emission from galaxy clusters}
\label{sub:Gamma-ray emission from galaxy clusters}
There is a fascinating possibility that GLAST finds clusters of galaxies
in $\gamma$-rays \citep{Berezinsky1997,Colafrancesco1998,Loeb2000,
Totani2000,Waxman2000,Keshet2003,Keshet2004a,Miniati2003}.
A fraction of the EGRB may be due to these clusters of galaxies.
While no convincing detection has been made so far from EGRET
\citep{Reimer2003}, there are a few reports of marginal evidence for
correlation between the position of clusters and the EGRET data
\citep{Kawasaki2002,Scharf2002}.  In fact, there are
many EGRET unidentified sources whose positions are coincident with
Abell clusters or high galaxy density regions, but the physical
association cannot be established because of low statistics and large
EGRET error circles.  It is expected that GLAST will give us the first
conclusive evidence for $\gamma$-ray emission from clusters.

Clusters of galaxies may emit $\gamma$-ray via two processes.
One is the collision between relativistic protons accelerated by shock
waves and surrounding cold matter, mainly protons, producing neutral
pions which decay into $\gamma$-rays.
Since protons hardly lose their energy by radiative loss and their
diffusion time is much longer than the age of the universe, all the
clusters are expected to emit some $\gamma$-rays by this mechanism
\citep{Berezinsky1997,Colafrancesco1998}.
The other mechanism is the inverse-Compton scattering of relativistic
electrons off the CMB photons \citep{Loeb2000,Totani2000,Waxman2000,
Keshet2003,Keshet2004a,Miniati2003}.
As for the source of these relativistic electrons, the most popular
scenario is that the shocks associated with the formation of large-scale
structure accelerate electrons to relativistic speed.
In the following sections we shall explore these two possibilities and
calculate the angular power spectrum that would be measured by GLAST.

\subsubsection{Proton-proton collisions}
\label{subsub:Proton-proton collisions}
The cumulative luminosity function, $\phi(z)$ [equation~(\ref{eq:phi
function})], for the cluster $\gamma$-ray emission from proton-proton
collisions is given by
\begin{equation}
 \phi_{C,pp}(z) = \int_{M_h(F_{\gamma,{\rm lim}},z)}^{\infty} dM_h\
  \frac{dn_h}{dM_h}(M_h,z),
  \label{eq:phi pp}
\end{equation}
where we label quantities regarding clusters by attaching the subscript
$C$ henceforth; another subscript $pp$ means that the $\gamma$-ray
emission comes originally from the proton-proton ($pp$) collisions.

In order to relate the $\gamma$-ray flux to the halo mass and redshift,
$M_h(F_\gamma,z)$, we follow the model of \citet{Colafrancesco1998}:
relativistic protons are injected from the very central region of
clusters, in which the central AGN or cD galaxy powers such an
injection.
These protons then diffuse from the central region to outside with the
efficiency that is determined by the magnetic field strength.
We assume that a fraction, $\epsilon_B = 10^{-3}$, of the baryon energy
is given to the magnetic energy.
For proton energies of our interest, the diffusion time scale is always
longer than the age of the universe; thus, protons are always confined
within clusters.
Using the radial injection profile of these diffused protons as well as
the density profile of the surrounding medium that is well measured in 
X-rays, one can compute the rate of $pp$ collisions.
The efficiency of the $\gamma$-ray production from each collision is
given in \citet{Kelner2006}, which we follow in our calculation.
We use a power law with an index of $\alpha_p = 2.2$ for the proton
spectrum, with an upper cutoff whose energy is determined by a balance
between the diffusion time scale and the cluster age
\citep{Colafrancesco1998}.
(We found that the cutoff energy is much larger than the energy scale
of our interest.)

\begin{figure}
\includegraphics[width=8.4cm]{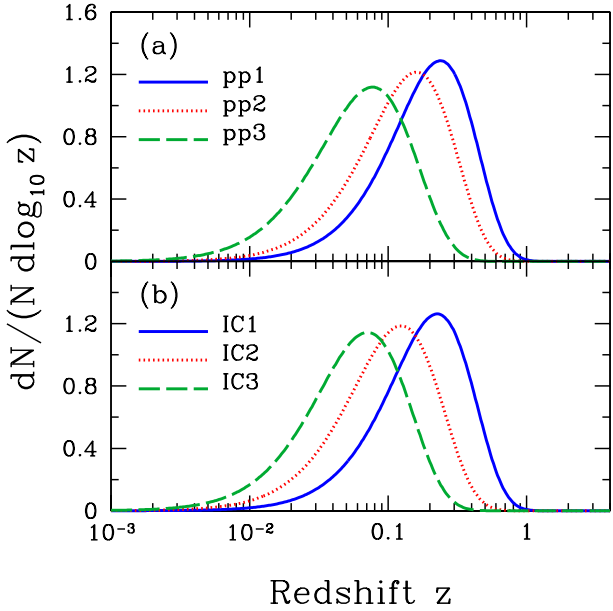}
\caption{Redshift distribution of clusters of galaxies that would be
 detected by GLAST for (a) proton-proton collision and (b)
 inverse-Compton scattering models. Model parameters are given in
 Table~\ref{table:cluster models}.}
\label{fig:zdist_cluster}
\end{figure}

We calculate the total energy of relativistic protons, $E_p$, by
assuming that a fraction, $\epsilon_p$, of the gravitational binding
energy of baryons is given to protons, i.e., $E_p = \epsilon_p E_b
\approx \epsilon_p (\Omega_b / \Omega_m) M V_c^2$, where $V_c$ is the
circular velocity at the virial radius, and adopt three values for
$\epsilon_p=0.5$ ($pp1$), 0.1 ($pp2$), and 0.01 ($pp3$).
Note that the equipartition model, $pp1$, has been excluded marginally
by observations \citep{Blasi1999}; however, we keep this model as an
upper bound on the $\gamma$-ray emission from clusters of galaxies via
proton-proton collisions.
The other two models ($pp2$ and $pp3$) are allowed by observations.
The most pessimistic cases, $pp3$ and IC3, are in agreement with
the estimates given in \citet{Gabici2004}.
We summarize the parameters of these models as well as the expected
number of clusters that would be detected by GLAST in
Table~\ref{table:cluster models}.
We find that a large number of galaxy clusters are expected to be seen
in the GLAST data, which would provide an exciting possibility of
investigating the physics of galaxy clusters using $\gamma$-ray
observations.
Figure~\ref{fig:zdist_cluster}(a) shows the redshift distribution of
clusters that would be detected by GLAST with these models.

\begin{table}
\caption{Model parameters, $\alpha$ and $\epsilon$, the expected number
 count, $N$, the surface density, ${\cal N}$, the average bias,
 $\overline b_C(z = 0.01)$, of clusters that would be  detected by
 GLAST. The last column lists the expected signal-to-noise ratio for
 detecting the correlation power spectrum averaged over $2\le l\le 30$.}
\label{table:cluster models}
\begin{minipage}{8.4cm}
\begin{tabular}{lcccccc}
 \hline
 Model & $\alpha_{p,e}$ & $\epsilon_{p,e}$ & $N$ &
 $\mathcal N$ (sr$^{-1}$) & $\overline b_C$ 
& $\overline C^C / \delta \overline C$\\
 \hline
 $pp1$ & 2.2 & 0.5 & 6600 & 530 & 2.0 & 7.0\\
 $pp2$ & 2.2 & 0.1 & 1100 & 88 & 2.5 & 4.6\\
 $pp3$ & 2.2 & 0.01 & 63 & 5.0 & 3.3 & 1.9\\
 IC1 & 2 & 0.05 & 3700 & 290 & 1.4 & 4.5\\
 IC2 & 2 & 0.01 & 430 & 34 & 1.7 & 2.4\\
 IC3 & 2.2 & 0.01 & 62 & 4.9 & 2.2 & 1.3\\
 \hline
\end{tabular}
\end{minipage}
\end{table}

\subsubsection{Inverse-Compton scattering}
\label{subsub:Inverse-Compton scattering}
Since electrons lose their kinetic energy via radiation loss rapidly
compared with the dynamical time of clusters of galaxies, $\gamma$-ray
emission would emerge only near the formation of shocks. 
The cumulative luminosity function, $\phi(z)$ [equation~(\ref{eq:phi function})], 
for the cluster $\gamma$-ray emission from the inverse-Compton
scattering is thus given by \citep{Totani2000}
\begin{equation}
 \phi_{C,{\rm IC}}(z) = \int_{M_h(F_{\gamma,{\rm lim}},z)}^{\infty} dM_h\
  R_{\rm form} (M_h,z)\Delta t_\gamma,
  \label{eq:phi IC}
\end{equation}
where $R_{\rm form}(M_h,z)$ is the formation rate of clusters with mass
of $M_h$ at $z$, per comoving volume, and $\Delta t_\gamma$ is the time
scale during which $\gamma$-rays are radiated efficiently from each
cluster.
We calculate $\Delta t_\gamma$ as either the inverse-Compton cooling
time or the shock wave propagation time (whichever is longer):
$\Delta t_\gamma = \max\{t_{\rm IC},t_{\rm shock}\}$.
In most cases of our interest, the latter is always much longer than the
former, and therefore, $\Delta t_\gamma = t_{\rm shock} \simeq r_{\rm
vir}/v_s = 1.5(1+z)^{-3/2}$ Gyr, independent of $M_h$, where $r_{\rm
vir}$ is the virial radius and $v_s$ the sound speed.
The formation rate of clusters, $R_{\rm form}(M_h,z)$, is given by the
time-derivative of the halo mass function, $dn_h/dM_h(M_h,z)$, corrected
for the halo destruction rate \citep{Kitayama1996}.

Similar to the proton-proton collision case, we calculate the total
energy of relativistic electrons, $E_e$, by assuming that a fraction,
$\epsilon_e$, of the gravitational binding energy of baryons is given to
electrons.
We use a power law with an index of $\alpha_e$ (either 2 or 2.2; see
Table~\ref{table:cluster models}) for the $\gamma$-ray spectrum, with an
upper cutoff whose energy is determined by a balance between the
acceleration time scale and the cooling time scale.
To calculate the acceleration time scale we use the magnetic field
energy given by $\epsilon_B = 10^{-3}$ times the binding energy of
baryons.
We choose $(\alpha_e, \epsilon_e) = (2,0.05), (2,0.01)$, and $(2.2,
0.01)$ as our models, and we call them  IC1, IC2, and IC3,
respectively.
The IC1 model is investigated by \citet{Totani2000}, and it gives
maximally allowed number of $\gamma$-ray emitting clusters, as the IC1
model predicts the EGRB flux that is as large as what is measured by
EGRET.
These models are again summarized in Table~\ref{table:cluster models}
and Fig.~\ref{fig:zdist_cluster}.

\subsection{Angular power spectrum of galaxy clusters from GLAST}
\label{sub:Results of cluster angular power spectrum}
The angular power spectrum of clusters of
galaxies\footnote{\citet{Waxman2000}, followed by
\citet{Keshet2003,Keshet2004a}, studied the angular correlation of the
radio and $\gamma$-ray background radiation from galaxy clusters.} is
given by
equations~(\ref{eq:C_l})--(\ref{eq:relation between w and xi}) with
the averaged blazar bias, $\overline b_B$, replaced by the average
cluster bias,
\begin{eqnarray}
 \overline b_{C,pp}(z) &=& \frac{1}{\phi_{C,pp}(z)}
  \int_{M_h(F_{\gamma,{\rm lim}},z)}^{\infty}
  dM_h\ \frac{dn_h}{dM_h}(M_h,z)
  \nonumber\\&&{}\times
  b_h(M_h,z),
  \label{eq:averaged bias pp}\\
 \overline b_{C,{\rm IC}}(z) &=& \frac{1}{\phi_{C,{\rm IC}}(z)}
  \int_{M_h(F_{\gamma,{\rm lim}},z)}^{\infty}
  dM_h\ R_{\rm form}(M_h,z) \Delta t_\gamma
  \nonumber\\&&{}\times
  b_h(M_h,z),
  \label{eq:averaged bias IC}
\end{eqnarray}
for the proton-proton collision model and the inverse-Compton model,
respectively.

\begin{figure*}
\begin{center}
\rotatebox{-90}{\includegraphics[width=12cm]{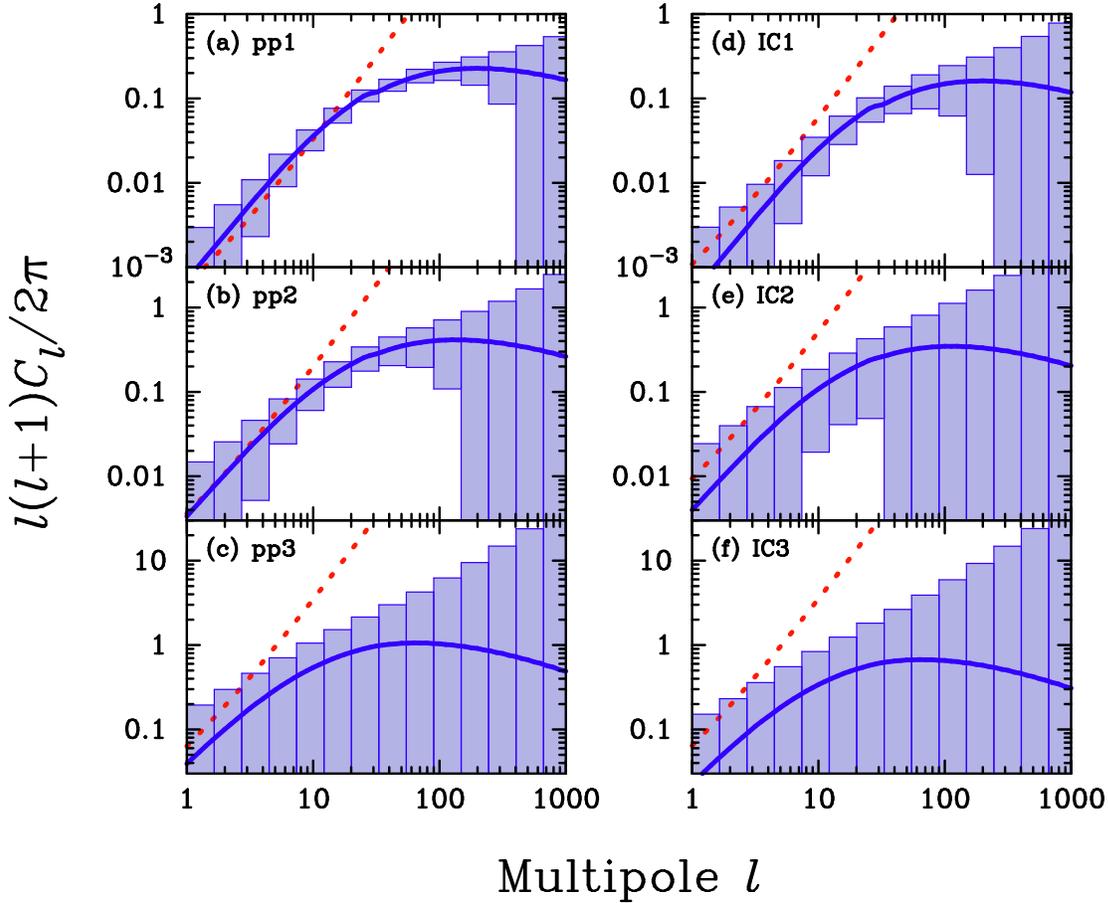}}
\caption{The same as Fig.~\ref{fig:C_l_blazar} but for galaxy
 clusters. The left panels show the proton-proton collision models
 ($pp1$, $pp2$, $pp3$), while the right panels show the inverse-Compton
 models (IC1, IC2, IC3). See Table~\ref{table:cluster models} for the
 model parameters.}
\label{fig:C_l_cluster}
\end{center}
\end{figure*}

Figure~\ref{fig:C_l_cluster} shows the angular power spectrum of these
$\gamma$-ray clusters with the binned error boxes ($\Delta l = 0.5 l$)
as well as the shot noise term for (a)--(c) proton-proton collision and
(d)--(f) inverse-Compton models.
The correlation is quite significant particularly for optimistic models
predicting large number of $\gamma$-ray emitting clusters being detected
by GLAST, i.e., $pp1$ and IC1. 
The last column of Table~\ref{table:cluster models} shows the
signal-to-noise ratio for the  power spectrum averaged over $2\le l\le
30$, $\overline C/\delta \overline C$.
We find that the signal-to-noise ratio exceeds unity for all the models
that we have considered: the minimum is $\overline C/\delta \overline
C=1.3$ for IC3, and the maximum is 7.0 for $pp1$, despite the fact that
only small number of clusters are expected to be seen in the GLAST
data.
This is because clusters of galaxies are formed in the high-density
peaks and thus are highly biased.
The sixth column of Table~\ref{table:cluster models} shows the average
bias factors of clusters at $z = 0.01$.


\section{Discussion}
\label{sec:Discussion}
\subsection{Admixture of blazars and galaxy clusters}
\label{sub:Source identification due to angular anisotropy}
While follow-up programs should reveal the identity of the GLAST
$\gamma$-ray sources and also some of the galaxy clusters might appear
as extended sources, at very early stage of GLAST observational
campaign, all the point sources should more generally be considered to
be mixed of various emitters.
Here we consider two-population case, blazars and galaxy clusters.
Our purpose in this section is to investigate whether it is possible to
distinguish the blazar component from that of clusters by the angular
clustering, even before the follow-ups.

When there are more than one species of sources on the sky, one should
also consider cross-correlation between different species.
When there are blazars and galaxy clusters in the $\gamma$-ray sky, the
angular power spectrum is given by
\begin{equation}
 C_l = C_{l,B} + C_{l,C} + 2 C_{l,BC},
  \label{eq:C_l blazar plus cluster}
\end{equation}
where $C_{l,B}$ and $C_{l,C}$ are the spectra from blazars
(Section~\ref{sec:Angular power spectrum of blazars}) and clusters
(Section~\ref{sec:Angular power spectrum of galaxy clusters}),
respectively.
The surface number density of sources in this case is instead given by
the sum of the two species, $\mathcal N = \mathcal N_B + \mathcal N_C$.
The last term, $C_{l,BC}$, represents the cross-correlation between
blazars and galaxy clusters, and is given by
\begin{eqnarray}
 C_{l,BC} &=&
  2\pi \int_{-1}^{1} d\cos\theta\
  P_l(\cos\theta) w_{BC}(\theta),
  \label{eq:angular power spectrum and cross correlation}\\
 \mathcal N^2 w_{BC}(\theta) &=&
  \int_0^{z_{\rm max}} dz\ \frac{d^2V}{dzd\Omega} \chi(z)^2
  \nonumber\\&&\times
  \phi_B(z)\phi_C(z)
  \overline b_B (z) \overline b_C (z)
  \nonumber\\&&\times
  \int_{-\infty}^{\infty}du\  \xi_{\rm lin}
  \left(\sqrt{u^2+\chi(z)^2\theta^2},z\right),
  \label{eq:cross correlation}
\end{eqnarray}
where $\phi_B(z)$ [equation~(\ref{eq:phi function})] and $\phi_C(z)$
[equation~(\ref{eq:phi pp}) or (\ref{eq:phi IC})] as well as $\overline
b_B(z)$ [equation~(\ref{eq:averaged bias})] and $\overline b_C(z)$
[equation~(\ref{eq:averaged bias pp}) or (\ref{eq:averaged bias IC})] have
been given in the previous sections.

\begin{figure*}
\begin{center}
\rotatebox{-90}{\includegraphics[width=12cm]{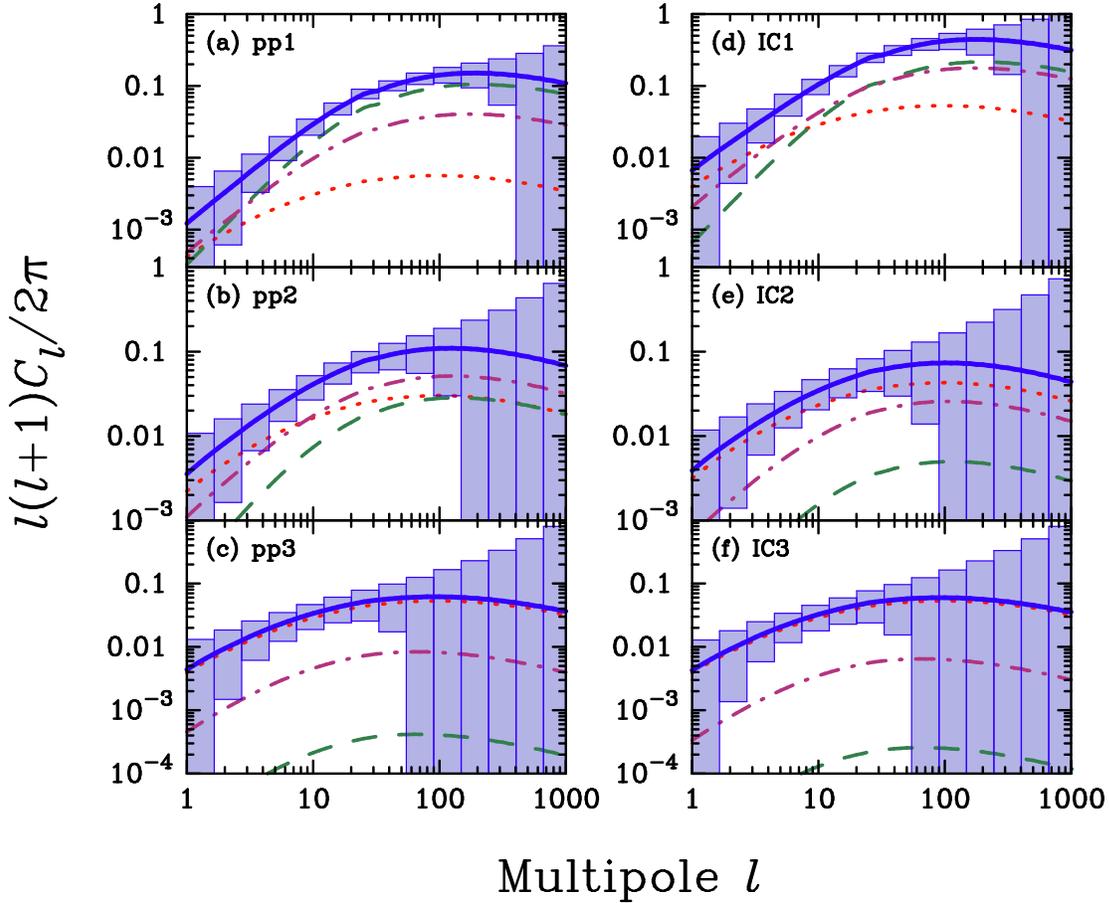}}
\caption{Angular power spectrum of blazars for the LDDE1 model with
 $\overline b_B = 3$ (dotted), clusters of galaxies (dashed), and
 cross-correlation between blazars and clusters (dot-dashed). The thick
 solid curves show the total signal, $C_l^C = C_{l,B}^C + C_{l,C}^C + 2
 C_{l,BC}$, while the boxes show the binned errors ($\Delta
 l=0.5l$). The left panels show the proton-proton collision models
 ($pp1$, $pp2$, $pp3$), while the right panels show the inverse-Compton
 models (IC1, IC2, IC3). See Table~\ref{table:cluster models}  for the
 model parameters.}
\label{fig:C_l_cross}
\end{center}
\end{figure*}

Figure~\ref{fig:C_l_cross} shows the angular power spectrum of GLAST
point sources, including blazars (LDDE1 and $\overline b_B = 3$) and
galaxy clusters for various models.
If the average blazar bias is as large as 3, the power spectrum from
blazars almost always dominates the signal, especially at low $l$'s.
An interesting feature in Fig.~\ref{fig:C_l_cross} is that the shape
of blazar power spectrum and cluster spectrum are quite distinct: the
cluster spectrum falling towards low multipoles more rapidly than the
blazar spectrum (i.e., the cluster spectrum rising towards high
multipoles more rapidly than the blazar spectrum).
Therefore, when the blazar signal dominates ($pp3$ and IC3 in
Fig.~\ref{fig:C_l_cross}), one would see a shallower power spectrum,
while when the cluster signal dominates ($pp1$ and IC1), one would see a
steeper power spectrum at $l\lesssim 30$.
This feature may help us identify the dominant source of clustering seen
by GLAST.
We do not know which point sources GLAST would detect, blazars or
clusters, until the follow-up observations are carried out; however, the
angular power spectrum may provide us with useful information regarding
the dominant species.

\subsection{Identifying GLAST point sources with radio survey}
\label{sub:Source identification due to radio survey}
The Faint Images of the Radio Sky at Twenty Centimeters (FIRST) survey
\citep{Becker1995} has detected $\sim 811$k radio sources over 9,033
square degrees, or $\sim 90$ sources per square degree, with the source
detection threshold of 1~mJy at
1.4~GHz.\footnote{http://sundog.stsci.edu/first}
Since many of the GLAST point sources will be within the FIRST field of
view, the FIRST survey will provide us with valuable information
regarding identification of the point sources that would be detected by
GLAST.

How bright are blazars and galaxy clusters in radio?
The EGRET blazar catalogue was constructed such that the EGRET sources
are also detected in radio.
The correlation between radio and $\gamma$-ray luminosities of blazars
is not yet well understood, and there is a considerable dispersion
\citep{Mucke1997}.
However, the standard synchrotron self-Compton model of blazars predicts
that there should be some correlation.
\citet{Narumoto2006} assumed a proportional relation between mean radio
luminosity, $L_r$, and $L_\gamma$, with a dispersion obeying to the
log-normal distribution, and found that mean relation of $L_r =
10^{-3.23} L_\gamma$ and dispersion $\sigma(\log(L_r/L_\gamma)) = 0.49$
fitted the observed data well.
Using this relation and assuming that the spectral index in radio is
given by $\alpha_r = 1.0$ (because the radio emission is due to
synchrotron radiation), we find that the flux in radio that corresponds
to the limiting flux for the point sources that can be detected by GLAST
in $\gamma$-rays is given by $F_{r,{\rm GLAST,lim}} = (1+z)^{2 -
\alpha_r} (10^{-3.23} L_\gamma)/(4\pi d_L^2) = 10^{-3.23}
(1+z)^{\alpha_\gamma - \alpha_r}(\alpha_\gamma - 1) E_{\rm min}
F_{\gamma,{\rm lim}}$, where we have used equation~(\ref{eq:limflux}).
We thus find $F_{r,{\rm GLAST,lim}} \sim 10$~mJy, which is an order of
magnitude brighter than the limiting flux for the point sources detected
by the FIRST survey.\footnote{The luminosities at 1.4 GHz and at 2.7 GHz
are the same, as we adopt $\alpha_r = 1$.}
Therefore, we expect the radio counterparts for the GLAST blazars to be
found in the existing FIRST point source catalogue (if they are within
the FIRST field of view), although some sources which deviate from the
$L_r$--$L_\gamma$ relation above may be missed.

The radio emission from galaxy clusters is also likely from synchrotron
radiation.
For the inverse-Compton scenario one can estimate the luminosity in
radio from a ratio of the CMB and magnetic field energy density, in
which case the radio luminosity from galaxy clusters is much smaller
than that expected from blazars having the same $\gamma$-ray luminosity
\citep{Totani2000}.
For the proton-proton collision scenario they would be even dimmer in
radio; otherwise they should also be detectable by $\gamma$-rays due
to the electron inverse-Compton scattering.
Therefore, unlike blazars, galaxy clusters would not be identifiable
with the FIRST survey, which makes the FIRST survey a good diagnosis
tool for identification of the GLAST point sources.


\subsection{Measuring blazar anisotropy  with GLAST}
\label{sub:Proposal for blazar anisotropy measurement with GLAST}
Based upon the results that we have obtained so far, we here show one
example strategy for the point source survey and identification of
blazars that would be carried out by GLAST.

(i) {\it Source detection.---}%
After its launch, GLAST will start detecting $\gamma$-ray sources from
all the directions on the sky.
Some of them would be extended sources (such as nearby galaxy clusters),
and some would be highly variable (such as $\gamma$-ray bursts).
These sources should be removed.

(ii) {\it Removing galaxy clusters.---}%
As blazars are also bright in radio but galaxy clusters are not, one may
remove galaxy clusters from the GLAST data using the source catalogue
from the FIRST survey (Section~\ref{sub:Source identification due to
radio survey}).

(iii) {\it Updating GLF and further cut.---}%
With the GLAST source catalogue from (ii), which would consist mostly of
blazars, one may update the GLF of blazars by extending it down to
fainter sources than those that EGRET has detected.
At this point we probably gain some insight as to which GLF model fits
the data better, LDDE1 or LDDE2, or whether or not we need a different
GLF model.
If LDDE1 is indeed confirmed, then one needs other populations of
sources in order to explain the bulk of EGRB.

(iv) {\it Analysis of angular power spectrum.---}%
Measure the angular power spectrum of the sources that have survived the
cuts in (i) and (ii).
Since we have currently several models which predict a variety of the
blazar bias, from 0.4 to 4 (Section~\ref{sec:Modeling blazar bias}), the
power spectrum measured at $l \lesssim 30$ should provide us with useful
information regarding the blazar bias.
While we would expect the contribution from galaxy clusters is minimal
in this catalogue owing to the cuts in (i) and (ii), the shape of the
power spectrum would also provide useful (albeit indirect) confirmation
that the bulk of the sources in the catalogue are blazars.
The blazar bias measured from GLAST, or an upper limit on the bias,
would be the first direct measurement of the bias of blazars, which
would shed light on the formation process of blazars and their link to
the quasars detected in the optical and the AGNs detected in X-rays.

(v) {\it Completion of follow-ups: beginning of precision study.---}%
When the source identification with direct follow-up observations is
complete, one should revisit the blazar source catalogue again,
establish the GLF of blazars more firmly, and re-analyze the angular
power spectrum.
At this point it would also be possible to obtain the 3-d power
spectrum, as opposed to the angular spectrum, using the redshift
information from follow-up observations.
This would be very powerful in constraining the formation and evolution
of blazars, as one can constrain the evolution of blazar bias as a
function of redshift, provided that enough number of blazars are
detected by GLAST.

\section{Conclusions}
\label{sec:Conclusions}
In this paper we have calculated the angular power spectrum of blazars
and galaxy clusters that would be detected by GLAST.
We have shown that GLAST can detect the spatial clustering of blazars if
the average bias of blazars exceeds 1.2 and 0.5, for the canonical GLF
model (LDDE1) that accounts for 25--50\% of the extragalactic
$\gamma$-ray background (EGRB) and the extreme model (LDDE2) that
accounts for all the EGRB, respectively \citep{Narumoto2006}.
While the blazar bias is not known with any precision, current
observations seem to suggest, albeit indirectly, that it can take on any
values between $\sim 0.4$ and $\sim 4$; thus, the GLAST data will
provide us with the first, direct estimate of the bias of blazars which,
in turn, would constrain the formation and evolution of blazars. 

As for galaxy clusters, which are highly biased objects, we have found
that the signal-to-noise ratio of the correlation exceeds unity for all
the models we have considered (Table~\ref{table:cluster models}):
proton-proton collisions followed by pion decay \citep{Berezinsky1997,
Colafrancesco1998}, and inverse-Compton scattering of relativistic
electrons off CMB \citep{Loeb2000,Totani2000,Waxman2000,Keshet2003,Keshet2004a,Miniati2003}.

We have shown that the angular power spectra of blazars and galaxy
clusters are quite distinct at low multipoles, $l\lesssim 30$, the
blazar spectrum being much shallower than the cluster one.
This feature helps us identify the population dominating the angular
power spectrum of the point sources that would be detected by GLAST.

Although the full follow-up observations would take long time, a quick
(but less accurate) identification of sources is possible with the
existing FIRST survey data in radio at 1.4~GHz, as most of the blazars
should also be bright enough to be seen in radio, while galaxy clusters
should not.
We have given an example strategy for using the angular power spectrum
as a diagnosis tool for blazars in Section~\ref{sub:Proposal
for blazar anisotropy measurement with GLAST}.

With an impressive number of blazars as well as galaxy clusters expected
to be detected by GLAST, one should maximize the scientific outcome from
the GLAST data by using as many tools as possible.
The angular power spectrum (or angular correlation function) is easy to
calculate from the point source catalogue, and would give us invaluable
information about the spatial clustering of blazars and high-energy
activity in clusters of galaxies, which are poorly known at present.

\section*{Acknowledgments}
S.A. and E.K. would like to thank Jennifer Carson for useful
discussion, and S.A. is also grateful to Andrew Benson and Marc
Kamionkowski for comments.
S.A. was supported by Sherman Fairchild Fellowship at Caltech.
E.K. acknowledges support from an Alfred P. Sloan Foundation.
T.N. and T.T. were supported by a Grant-in-Aid for the 21st Century
COE ``Center for Diversity and Universality in Physics'' from the
Ministry of Education, Culture, Sports, Science and Technology of
Japan.

\label{lastpage}

\end{document}